\documentstyle{mn}

\newif\ifAMStwofonts

\ifoldfss
  \ifCUPmtlplainloaded \else
    \NewTextAlphabet{textbfit} {cmbxti10} {}
    \NewTextAlphabet{textbfss} {cmssbx10} {}
    \NewMathAlphabet{mathbfit} {cmbxti10} {} 
    \NewMathAlphabet{mathbfss} {cmssbx10} {} 
  \fi
  \ifAMStwofonts
    \ifCUPmtlplainloaded \else
      \NewSymbolFont{upmath} {eurm10}
      \NewSymbolFont{AMSa} {msam10}
      \NewMathSymbol{\upi}     {0}{upmath}{19}
      \NewMathSymbol{\umu}     {0}{upmath}{16}
      \NewMathSymbol{\upartial}{0}{upmath}{40}
      \NewMathSymbol{\leqslant}{3}{AMSa}{36}
      \NewMathSymbol{\geqslant}{3}{AMSa}{3E}

       \let\le=\leqslant
       \let\ge=\geqslant
    \fi
  \fi
\fi 

\ifnfssone
  \newmathalphabet{\mathit}
  \addtoversion{normal}{\mathit}{cmr}{m}{it}
  \addtoversion{bold}{\mathit}{cmr}{bx}{it}
  \newmathalphabet{\mathbfit} 
  \addtoversion{normal}{\mathbfit}{cmr}{bx}{it}
  \addtoversion{bold}{\mathbfit}{cmr}{bx}{it}
  \newmathalphabet{\mathbfss} 
  \addtoversion{normal}{\mathbfss}{cmss}{bx}{n}
  \addtoversion{bold}{\mathbfss}{cmss}{bx}{n}
  \ifAMStwofonts
    \ifCUPmtlplainloaded \else
      \UseAMStwoboldmath
      \makeatletter
      \new@mathgroup\upmath@group
      \define@mathgroup\mv@normal\upmath@group{eur}{m}{n}
      \define@mathgroup\mv@bold\upmath@group{eur}{b}{n}
      \edef\UPM{\hexnumber\upmath@group}
      \new@mathgroup\amsa@group
      \define@mathgroup\mv@normal\amsa@group{msa}{m}{n}
      \define@mathgroup\mv@bold\amsa@group{msa}{m}{n}
      \edef\AMSa{\hexnumber\amsa@group}
      \makeatother
      \mathchardef\upi="0\UPM19
      \mathchardef\umu="0\UPM16
      \mathchardef\upartial="0\UPM40
      \mathchardef\leqslant="3\AMSa36
      \mathchardef\geqslant="3\AMSa3E

       \let\le=\leqslant
       \let\ge=\geqslant
    \fi
  \fi
\fi 

\ifnfsstwo
  \DeclareMathAlphabet{\mathbfit}{OT1}{cmr}{bx}{it}
  \SetMathAlphabet\mathbfit{bold}{OT1}{cmr}{bx}{it}
  \DeclareMathAlphabet{\mathbfss}{OT1}{cmss}{bx}{n}
  \SetMathAlphabet\mathbfss{bold}{OT1}{cmss}{bx}{n}
  \ifAMStwofonts
    \ifCUPmtlplainloaded \else
      \DeclareSymbolFont{UPM}{U}{eur}{m}{n}
      \SetSymbolFont{UPM}{bold}{U}{eur}{b}{n}
      \DeclareSymbolFont{AMSa}{U}{msa}{m}{n}
      \DeclareMathSymbol{\upi}{0}{UPM}{"19}
      \DeclareMathSymbol{\umu}{0}{UPM}{"16}
      \DeclareMathSymbol{\upartial}{0}{UPM}{"40}
      \DeclareMathSymbol{\leqslant}{3}{AMSa}{"36}
      \DeclareMathSymbol{\geqslant}{3}{AMSa}{"3E}

       \let\le=\leqslant
       \let\ge=\geqslant
    \fi
  \fi
\fi 

\ifCUPmtlplainloaded \else
  \ifAMStwofonts \else 
    \def\upi{\pi}
    \def\umu{\mu}
    \def\upartial{\partial}
  \fi
\fi

\title{Long-Term Variability of the Low-Luminosity Active Galactic Nucleus of M81}
\author[N. Iyomoto]
       {Naoko Iyomoto$^1$, Kazuo Makishima$^2$\\
	1. The Institute of Space and Astronautical Science,
	3-1-1 Yoshinodai, Sagamihara, Kanagawa 229-8510, Japan\\
        2. Department of Physics, School of Science, 
	The University of Tokyo, 7-3-1 Hongo, Bunkyo-ku, Tokyo 113-0033, Japan}
\date{}
\date{Accepted 2000 September 22.
      Received 2000 August 14;
      in original form 2000 April 28.}

\pubyear{2000}

\begin{document}
\maketitle
\begin{abstract}
Long-term X-ray variability of 
the low-luminosity active galactic nucleus of M81 was studied,
using 16 {\it ASCA} observations spanning 5.5 years.
The object exhibits a factor three variation over the 5.5 years.
The source intensity was relatively constant within each observation
which lasted typically for one day,
but intra-day variability by 30\% was detected on the 15th observation.
The power-spectral density (PSD) was estimated in ``forward'' manner,
over a frequency range of $10^{-8.2}$--$10^{-4.3}$ Hz 
(period range of 1/4 day--5.5 years), 
by utilizing structure function and extensive Monte-Carlo simulations
in order to overcome the very sparse and uneven data samplings.
When the PSD is assumed to be white below a ``break frequency'' $f_{\rm b}$
and falls off as $\propto f^{-\alpha}$ above $f_{\rm b}$,
where $f$ is frequency and $\alpha$ is a positive parameter, 
the M81 light curve is well described with 
1/$f_{\rm b} \ge 800$ days and $\alpha = 1.4 \pm 0.2$.
\end{abstract}

\noindent {\bf Key words}\\
galaxies: active -- galaxies: individual: M81 (NGC~3031) -- galaxies: nuclei -- methods: data analysis -- methods: numerical -- X-rays: galaxies --

\section{INTRODUCTION}
A number of low-luminosity AGNs (LLAGNs), 
with X-ray luminosities in the range $10^{40-42}$ erg s$^{-1}$,
have been discovered in nearby normal galaxies with {\it ASCA}
(e.g. Makishima et al. 1994; Iyomoto et al. 1996; Ishisaki et al. 1996;
Iyomoto et al. 1998; Terashima et al. 1998; Ptak et al. 1999).
LLAGNs are now considered to be ubiquitous at least in our neighbourhood.

The low luminosities of LLAGNs imply lower black-hole masses 
and/or lower mass accretion rates 
as compared to more luminous AGNs.
These alternatives may not be distinguished by X-ray spectra,
since AGNs exhibit similar X-ray spectra over wide range of luminosities,
from LLAGNs through quasars.
On the other hand, 
X-ray variability is expected to provide useful clues to the issue,
as a zero-th order approximation, 
because characteristic time scales of the X-ray emission is considered to 
correlate positively with the system size and hence the system mass.

During individual observations with {\it ASCA},
which typically last for one day,
LLAGNs tend to be much less variable than 
the lower-luminosity end of the Seyfert galaxies (e.g. Ptak et al. 1998).
We are therefore tempted to speculate that 
LLAGNs have longer characteristic time scales of the X-ray variation
and hence larger black-hole masses compared to Seyferts.
In this case, LLAGNs might be ``dead quasars'' in our neighbourhood.
To examine this possibility,
it is important to determine the power spectral density (PSD) 
of intensity variations of LLAGNs 
on time scales such as dozens of days or longer.
So far, no such attempts have been made,
except for a rough estimation by Ishisaki et al. (1996),
because only few observations with uneven intervals
are usually available for each LLAGN in X-rays.

Using {\it ASCA},
we investigate long-term variability of an LLAGN at the nucleus of M81,
which is an Sab galaxy at a distance of 3.6 Mpc (Freedman et al. 1994).
Its nucleus exhibits weak AGN properties in various wave bands.
Optically, it is classified as a LINER (Ho et al. 1997).
In radio, the spectrum from the nuclear region 
is interpreted as a jet emission (Boettcher et al. 1997).
The nucleus is also an X-ray source with a 2--10 keV luminosity
of $\sim 2 \times 10^{40}$ erg s$^{-1}$, as confirmed repeatedly by 
{\it Einstein Observatory} (Fabbiano et al. 1988),
{\it EXOSAT} (Barr et al. 1995),
{\it Ginga} (Tsuru 1992),
{\it ROSAT} (Boller et al. 1992), 
{\it BBXRT} (Petre et al. 1993),
{\it ASCA} (Ishisaki et al. 1996),
and {\it BeppoSAX} (Pellegrini et al. 2000).
These multi-mission results indicate a mild X-ray intensity variation.
Ishisaki et al. (1996),
using multiple observations with {\it ASCA} spanning $\sim$2.5 years,
detected long-term variation 
as well as a short-term ($\sim 1$ day) variation on one occasion.
Pellegrini et al. (2000) detected
a periodic oscillation with a sub-day period.
In addition, the spectra of this LLAGN,
except that measured with {\it Einstein},
are well fitted with a power-law model having a photon index of 1.5--2.0.
Therefore, M81 has been considered to possess an LLAGN.

\section{OBSERVATIONS AND DATA REDUCTION}
Owing to the explosion of the supernova SN 1993J (Kohmura et al. 1994) in M81, 
this particular field has been observed with {\it ASCA} 
16 times over 5.5 years,
as summarized in the observation log of Table~1.
It is among the objects most frequently observed with {\it ASCA}.
The results on SN 1993J have been reported
by Kohmura (1994), Kohmura et al. (1994), and Uno (1997).
The results on the M81 nucleus of the 1st to 10th observations
have been published as Ishisaki et al. (1996).

The 16 observations were carried out with 
the GIS (Ohashi et al. 1996; Makishima et al. 1996) in the PH mode,
while the SIS (Burke et al. 1991) in several different modes 
as described in Table~1.
The M81 nucleus typically gave 
an intensity of 0.3 c s$^{-1}$ for the GIS (0.7--10 keV, for each detector)
and 0.4 c s$^{-1}$ for the SIS (0.5--10 keV, for each detector).
According to Kohmura et al. (1994),
the SIS count rate of SN 1993J was $\sim$ 25\% of that of the nucleus
in the 1st observation,
and decreased as $t^{-0.7}$,
where $t$ is the time after the explosion.

For the first six observations when SN~1993J was bright,
we accumulated photons $1'.5$ around the nucleus
to avoid the contamination from SN~1993J.
Since SN~1993J exhibit hard spectrum in this epoch (Kohmura et al. 1994),
we excluded photons above 4 keV in the following analysis.
For the other 10 observations, 
we accumulated photons $3'$ around the nucleus,
and considered all photons in the analysis.
The SN~1993J contribution to the data accumulation region
is expected to be $< 5$\%.

For the spectral analysis,
we accumulated background spectra using the blank-sky data provided by NASA.
The background photons are $< 2$\% of the signal photons.
For the timing analysis,
we made background light curve 
using the blank regions of the M81 field.

\section{SPECTRA AND LIGHT CURVES}
\label{sec:spec}
Figure~\ref{fig:longLC} shows long-term light curves of the M81 nucleus,
where the 2--10 keV flux in each observation is obtained by spectral fitting,
in the following manner.
The spectra of the M81 nucleus are dominated 
by a featureless hard continuum 
without significant absorption or sharp line feature,
as Ishisaki et al. (1996) reported,
so that
we adopted single power-law model.
We fixed photon index $\Gamma$ at 1.8
and determined the fluxes in Figure~\ref{fig:longLC},
since $\Gamma$ has remained almost constant around 1.8
over 5.5 years.
The GIS and SIS give consistent fluxes within systematic errors.
The results on the 5th through 10th observations
agree with those reported by Ishisaki et al. (1996).
Through the 1st to 6th observations conducted during the first 1.5 months,
the intensity remained almost constant.
On the other hand, 
over the other 10 observations performed every six months,
peak-to-peak intensity variation amounts to a factor of three.

We also analyzed short-term variability of the source intensity
as shown in Figure~\ref{fig:shortLC},
by subdividing each observation into time bins of 1/8-day width.
With the 1/8-day bin, 
typical error due to photon counting statistics
becomes $6\times10^{-13}$ erg cm$^{-2}$ s$^{-1}$,
or $\sim$3\% of the average flux.

We summed up the GIS and SIS counts,
since the two instruments give consistent results.
Considering the integration radius and positional response
of individual observations and the spectral information,
we converted the count rates into 2--10 keV fluxes.
Before the conversion,
we subtracted 
the background and the contamination from SN~1993J,
using the average value of each observation.
We fitted the light curves assuming a constant count rate,
to evaluate the intensity variation through the $\chi^2$ test.
The obtained values of $\chi^2$ and degree-of-freedom (d.o.f.)
are shown in the left bottom of the figures.
Thus, the source exhibits statistically significant short-term variation
in the 5th, 6th, 8th, 11th, 12th, 14th and 15th observations,
in addition to the 3rd (Serlemitsos et al. 1996)
and 9th (Ishisaki et al. 1996) observations, as already reported.

\section{TIMING ANALYSIS}
\label{sec:time}
With the 1/8-day bin width, there are 112 light-curve bins in total,
of which the average flux becomes
$2 \times 10^{-11}$ erg cm$^{-2}$ s$^{-1}$.
Root-mean-square (RMS) of the variation becomes 
$7 \times 10^{-12}$ erg cm$^{-2}$ s$^{-1}$,
an order of magnitude larger than Poisson error.
As already attempted by Ishisaki et al. (1996),
we may estimate a gross continuum shape of the M81 PSD
using the 112-point light curve,
because it includes various sampling intervals: hours, days, months and years.

However, the light curve is subject to an extremely uneven and sparse sampling.
To illustrate it,
if M81 had been observed continuously
during the total observed span (2022 days),
we would obtain $2022 \times 8$=16176 bins with the 1/8-day bin width. 
In practice, there are only 112 bins.
More than 99\% of the total span are thus data gaps.
Consequently, it is difficult to
determine the M81 PSD in general manner, i.e.
Fourier transforming the light curve into ``periodogram'',
and obtaining a PSD as modulus squared of the periodogram.

We have introduced two key methods 
to estimate the long-term PSD from the sparse light curve
(Iyomoto 1999; Iyomoto et al. 1999).
The first is ``forward method''.
Namely, assuming various PSDs, 
we generate fake light curves and compare them with the actual M81 data
to estimate which model PSD can best reproduce the actual M81 data.
We utilized Monte-Carlo technique
to evaluate the effects of the data sampling gaps. 
The second is the use of "structure function", hereafter SF,
which is mathematically equivalent to PSD,
but is less affected by data gaps.
We convert the light curves into SFs,
before comparing the actual data and the simulation,
because it is difficult to compare light curves directly.

In \S 4.1, we calculate the M81 SF.
In \S 4.2, assuming various PSDs, 
we generate fake SFs via Monte-Carlo technique.
In \S 4.3, we compare the fake and actual SFs
to determine break frequency, $f_{\rm b}$, and slope index, $\alpha$, of PSD.
In \S 4.4, we determine the normalization of PSD, $P_0$,
directly from the RMS of the M81 light curve,
and derive confidence regions of the M81 PSD.

\subsection{Structure Function}
\label{subsec:sf}
The definition of first-order structure function
is given by Simonetti et al. (1985) as
\begin{equation}
S(\tau_k) =
\frac{1}{N(\tau_k)}\sum_{m=0}^{N-1-n} \left(l_m-l_{n+m} \right)^2~~,
\label{eqn:SF}
\end{equation}
where $l_n$ is luminosity at time $t_n$.
The sum is taken over data pairs satisfying $t_{n+m}-t_m$=$\tau_k$,
with $\tau_k$ ($k$=1,2,...) being the time-lag,
and $N(\tau_k)$ denotes the number of such pairs.
For a stationary random process with equally spaced sampling,
SF, auto-correlation-function (ACF),
and PSD are equivalent to one another.
For example, SF is related to ACF, $A(\tau_k)$, as
\begin{equation} 
S(\tau_k) = 2\sigma^2[1-A(\tau_k)]~~,
\label{eqn5:ACFtoSF}
\end{equation}
where $\sigma^2$ is variance of the process.

Nevertheless, SF is much less sensitive to data gaps
compared to ACF and periodogram, and hence to PSD (e.g. Hughes et al. 1992).
In astronomy, SF have been used in various research fields
including radio (e.g. Hughes et al. 1992), 
optical (e.g. Paltani et al. 1997, Kawaguchi et al. 1998),
and X-ray (e.g. Leighly et al. 1997; Heidt et al. 1998) frequencies.

Figure~\ref{fig:SF}a shows the M81 SFs calculated by eq.~(\ref{eqn:SF}).
In addition to the SF derived from the 1/8-day bin light curve,
we superposed small-time-lag portion of SFs 
derived from 1/4, 1/16, 1/32 and 1/64-day bin light curves, for comparison.
The SFs with shorter light-curve bin,
especially those with 1/32 and 1/64-day binning,
are strongly contaminated by the Poisson noise.
On the other hand,
it is reasonable to suppose that
the Poisson noise becomes negligible 
for the SFs with light-curve binning longer than 1/8 day,
because the 1/4 and 1/8-day bin SFs are similar to one another.
It is consistent with the fact 
that the Poisson noise is $\sim$1/10 
of the RMS amplitude of the M81 variation in the 1/8-day bin light curve.
Accordingly, we utilize the 1/8-day binning in this paper.

Thanks to the various sampling intervals involved in our data,
we can calculate the SF over a wide frequency range.
SFs with time-lags shorter than one day
come from within individual observations.
SFs with time-lags between two days and 50 days are owing to 
various combinations among the 1st to the 6th observations.
SFs with time-lags longer than six months are
provided by combinations of observations
which are separated by multiples of six month.

For time-lags shorter than one day, the M81 SF is smooth.
Relatively large values of $N(\tau_k)$ may account for the smoothness.
On the contrary, in the larger time-lag region,
the calculated SF exhibits a large scatter, due to small $N(\tau_k)$.
Therefore, hereafter,
we average the SF into appropriate bins in time-lag,
as shown in Figure~\ref{fig:SF}b.
This is justifiable because we are interested in the PSD continuum,
rather than its fine structure.

The binned M81 SF seems to keep increasing
towards the longer end of the time coverage.
In other words, 
it is likely that the characteristic time scale of the M81 nucleus 
is longer than the total observed span of 2022 days.
However, to make this inferences more secure,
we need a more quantitative evaluation.
Even the binned SF is subject to strong sampling bias, which 
depends on the sampling window.
Consequently, it is difficult to directly convert the M81 SF into a PSD.
Similarly, it is difficult to directly compare SFs
suffering from different sampling windows.
Instead, we here take a forward-method approach to find out a set of PSDs
that are consistent with the observed SF.
That is, we start from an assumed PSD,
produce a number of Monte-Carlo light curves 
based on the assumed PSD and employing the actual sampling window,
and convert each fake light curve into a fake SF.
Then, we can study how the ensemble of SFs derived from a single PSD
behaves statistically under the particular sampling window,
and whether the actual M81 SF can be considered 
as a member of this ensemble or not.
This is carried out in the following subsections.

\subsection{Monte-Carlo Simulation}
\label{subsec:Monte-Carlo}
In order to qualitatively study the characteristic time scale of the variation
and to estimate its lower limit, 
we hypothesized that 
the M81 PSD is expressed by a convex broken power-law as
\begin{equation}
P(f) = \left\{ \begin{array}{ll}
          P_0                        & (f < f_{\rm b}) \\
          P_0 \; (f/f_{\rm b})^{-\alpha} & (f > f_{\rm b})
         \end{array}
         \right.  ~~,
\label{eqn:PSD}
\end{equation}
as a start point of the forward method.
Here $f$ denotes the frequency, $P_0$ is a constant,
$f_{\rm b}$ is a characteristic frequency called ``break frequency'',
and $\alpha$ is a slope index.
It is an empirical PSD of AGNs:
white at low frequency and red at high frequency
(e.g. Pounds \& McHardy 1988; Edelson \& Nandra 1998; Hayashida et al. 1998).
In this paper, we use this formula
as a template to describe our result on M81,
and try to estimate the values of $f_{\rm b}$, $\alpha$ and $P_0$.
We examine in \S~\ref{subsec:chi2}
whether this hypothesis is appropriate or not.
We regard $f_{\rm b}$ and $\alpha$ as free parameters of the forward method.
The value of $P_0$ is adjusted in \S~\ref{subsec:norm}.
The trial values of 
1/$f_{\rm b}$ are $10^{1.0}$, $10^{1.2}$, $10^{1.4}$, ..., $10^{4.8}$ days,
and those of $\alpha$ are $1.10$, $1.15$, $1.20$, ..., $2.10$.
We consider Fourier periods from 
1/$f_1 = 65536$ days to 1/$f_2 = 1/4$ day,
so that there are 2$^{18}$(=262144) bins in the PSD.
We adopted the 1/$f_1$ that is $\sim 32$ times longer 
than the total observed span of the actual data,
considering that 1/$f_1$ of the actual PSD is infinite,
and that the Fourier components
longer than the total observed span also
contribute to the observed light curve.

Assuming a particular set of $f_{\rm b}$ and $\alpha$,
we generate fake SFs in the following steps.\\
(i) Using Monte-Carlo simulation, 
we generate a set, namely 2$^{18}$(=262144), of random numbers,
uniformly distributed between 0 and 2$\pi$,
and use them for the phase of each Fourier component of the assumed PSD.\\
(ii) By a Fourier transform of this Monte-Carlo generated periodogram,
we obtain a light curve of 2$^{19}$(=524288) bins,
with 1/8-day bin width and 65536-days total span.\\
(iii) From the 524288-point light curve,
we select 112 data points by applying the M81 sampling window.\\
(iv) We normalize the 112-point light curve
to have an RMS equal to that of the M81 light curve.\\
(v) We calculate SF from the 112-point light curve.\\
If we use another set of random numbers and go through the above steps,
we can generate a different light curve from the same PSD.
In this way, we generate 1000 fake SFs
for each set of $f_{\rm b}$ and $\alpha$.

Figure~\ref{fig:SF}c provides examples of fake SFs
that were produced from 
a single PSD having $1/f_{\rm b}=10^{4.0}$ day and $\alpha=1.4$;
the SFs with $1/f_{\rm b}=10^{4.0}$ day represents 
those having $1/f_{\rm b}$ enough longer than the total observed span.
On large-time-lag range,
the fake SF scatters as largely as the M81 SF due to the sparse sampling.
Even on the time-lag range smaller than one day,
there are significant scatters between individual fake SFs,
arising from the sparse sampling,
while each SF is smooth in such time-lag range.
These are explained by mutual dependences among the SFs at different time-lags.
Namely, the smoothness in the small time-lag range 
is not only because $N(\tau_k)$ is large,
but because there are mutual dependences among the SFs at different time-lags.

Using ensemble properties of the 1000 fake SFs,
we can reduce the effect of the sparse sampling.
The ensemble average 
becomes smooth, as examplified in Figure~\ref{fig:SF}d.
Moreover, using standard deviation of these fake SFs,
we can estimate how much the SF scatter under the sparse sampling.
In Figure~\ref{fig:SF}d, we show the standard deviation as error bars.
Thus, we compare each ensemble-averaged fake SF with the actual SF
considering the standard deviation,
to examine whether the particular PSD assumed is appropriate or not.

It is worth pointing out 
that the 1/$f_1$ of the assumed PSD really have an effect on the fake SFs.
Figure~\ref{fig:SF}e shows the ensemble averages of fake SFs
with 1/$f_1$ of 2$^{11}$, 2$^{12}$, ..., 2$^{18}$ days.
The fake SFs having 1/$f_1 > 2^{16}$ days well reproduce
``drops'' around one day in the actual M81 SF.
We confirmed that similar dependence on 1/$f_1$ is seen
in the fake SFs with shorter characteristic time scale, 1/$f_{\rm b}$,
such as 10 days, 100 days and 1000 days.
These effect of 1/$f_1$ depends on the sampling window.
Therefore, it is important to select 1/$f_1$ 
that is enough longer than the total observed span of the actual light curve.
In this paper, we adopted 1/$f_1$ of 2$^{16}$ day.

\subsection{Comparison between Fake and Actual Data}
\label{subsec:chi2}
In order to quantitatively compare the actual and fake SFs,
we utilized the standard $\chi^2$ evaluation technique.
We defined $\chi^2$ as
\begin{equation}
\chi^2 
= \sum_{\tau_k} \left\{ \frac{S^{\rm M81}_{\tau_k} -\overline{S^{\rm fake}_{\tau_k}}}
{\sigma^{\rm fake}_{\tau_k}}\right\}^2~~,
\label{eqn:chi2}
\end{equation}
where $S^{\rm M81}_{\tau_k}$ is the M81 SF at a time-lag $\tau_k$, 
while $\overline{S^{\rm fake}_{\tau_k}}$ and $\sigma^{\rm fake}_{\tau_k}$ are 
the ensemble average and standard deviation among the 1000 fake SFs.
In this paper, we use the term $\sigma^{\rm fake}_{\tau_k}$ 
to refer to the data scattering due to the sampling window,
and not to the Poisson statistical nature of the data.
We regard this $\chi^2$ as 
a measure of relative goodness of the assumed PSD. 
We then calculated $\chi^2$ against various $f_{\rm b}$ and $\alpha$
and obtained the minimum $\chi^2$, namely $\chi^2_{\rm min}$ of 25,
at $1/f_{\rm b}=10^{4.6}$ days and $\alpha=1.40$.
We can regard these particular set of parameters
as the best estimate for $f_{\rm b}$ and $\alpha$ of the M81 PSD,
as long as the PSD of eq.~(\ref{eqn:PSD}) is assumed.
We should notice that, however, 
we give weight to the lower-limit to 1/$f_{\rm b}$
rather than the best-fit value,
because the obtained best-fit value is much longer than 
the total observed span, and in such 1/$f_{\rm b}$ range, 
the $\chi^2_{\rm min}$ values at different 1/$f_{\rm b}$
are almost similar to one another, 
as seen in Figure~\ref{fig:cont}a.

The next step is to estimate 68.3\%, 90\% and 99\% confidence regions.
However, we cannot apply
general relation between confidence levels and 
the $\Delta \chi^2$/d.o.f. values,
such as $\Delta \chi^2$/d.o.f. = 2.71
for the single parameter 90\% confidence limits. 
It is because the d.o.f. is ambiguous in the present case,
due to the mutual dependences among the SFs at different time-lags.
This effect is particularly prominent in the small-time-lag range,
as mentioned in $\S$~\ref{subsec:Monte-Carlo}.

Instead,
we estimate the $\Delta \chi^2$ values
that correspond to 68.3\%, 90\% and 99\% confidence boundaries, 
using again the Monte-Carlo simulation.
Employing the M81 best parameters obtained above,
we generated another set of 5000 fake light curves and hence fake SFs;
we would like to emphasize that
the $\Delta \chi^2$ distribution, which is given below,
is almost independent of the assumed PSD parameters.
Of course, the random numbers of these fake light curves
are different from those used in $\S$~\ref{subsec:Monte-Carlo}.
For each of the newly produced fake SF,
we determined the optimum $f_{\rm b}$ and $\alpha$,
in exactly the same way as performed for the M81 data.
We also calculated $\chi^2$ associated with each simulation
at the assumed PSD parameters themselves:
$1/f_{\rm b}=10^{4.6}$ days and $\alpha=1.40$.
We denote it $\chi^2_{\rm true}$.
Note that
the optimum parameters do not always coincide with the assumed parameters,
and that
$\chi^2_{\rm min}$, calculated for the optimum parameters,
is equal to or smaller than $\chi^2_{\rm true}$.
For each fake SF, we then calculated
\begin{equation}
\Delta \chi^2 \equiv \chi^2_{\rm true} - \chi^2_{\rm min}
\end{equation}
which can be regarded as a measure of statistical fluctuation
associated with eq.~(\ref{eqn:chi2}).
Figure~\ref{fig:chi2}a 
shows the $\Delta \chi^2$ distribution 
of the 5000 simulations.
As can be seen there,
4500 out of 5000 simulations have $\Delta \chi^2 \le 23$.
Hence, we regard 
$\Delta \chi^2$ = 23 as definition of the 90\% confidence boundary.
Similarly, 
$\Delta \chi^2$ = 3 and $\Delta \chi^2$ = 55 
provides the 68.3\% and 99\% confidence boundary, respectively.

Now that we have determined the confidence limits
associated with the $\chi^2$ of eq.~(\ref{eqn:chi2}),
we return to the M81 SF and the original set of 1000 fake SFs,
and examine how much errors are allowed.
The two-parameter confidence ranges determined in this way
is shown in Figure~\ref{fig:cont}a on the plane of 1/$f_{\rm b}$ and $\alpha$.
The 90\% confidence regions of the characteristic time scale and the PSD index 
are determined to be 1/$f_{\rm b} \ge 800$ days and $\alpha = 1.4 \pm 0.2$,
respectively.
Our detailed analysis has revealed 
longer 1/$f_{\rm b}$
compared to the rough estimation of Ishisaki et al. (1996):
they considered 
RMS of a short-term light curve (the 9th observation)
and a long-term light curve (the 1st through 10th observations)
and assumed $\alpha =1.40$
to obtain $f_{\rm b}$=(1--3)$\times 10^{-6}$ Hz
(1/$f_{\rm b}$ = 4--12 days).

Now we can also examine
whether the hypothesized shape of PSD is appropriate or not,
using the $\chi^2_{\rm true}$ distribution of the 5000 simulations.
Generally, a reasonable fit gives a $\chi^2$ value similar to d.o.f.
However, as mentioned in \S~\ref{subsec:Monte-Carlo},
d.o.f. is not well determined in the present case.
Instead, we compared the $\chi^2_{\rm min}$ value of M81 data
with the $\chi^2_{\rm true}$ distribution of fake data,
as a d.o.f. independent estimation of the absolute goodness.
The $\chi^2_{\rm true}$ distribution of the fake light curves 
is shown in Figure~\ref{fig:chi2}b;
again, the distribution is 
nearly independent of the assumed PSD parameters of the 5000 fake SFs.
Because the $\chi^2_{\rm min}$ value of the actual M81 data, 
shown as an arrow,
lies within 90\% regions of the distribution,
we conclude that the assumed two-component PSD is appropriate
within the limitation of the current observation.
Similarly, the single power-law PSD also turned out to be appropriate.

We admit that, however, it dose not mean 
that we can exclude other shapes of PSD
which we have not yet examined.
Especially,
the low frequency portion of the M81 SF has such large scatters
that other types of PSD might explain the observed data
equally well, or even better.
For example, it may have a negative or positive index in low frequencies,
while we assumed zero index (white noise).
A more complex PSD,
for example three-component PSD seen in 
low-mass X-ray binaries (e.g. Wijnands et al. 1999)
and Galactic black holes (e.b. Miyamoto et al. 1994),
which consists of a red-noise part at low frequency,
a white-noise part at intermediate frequency,
and another red-noise part at high frequency,
might explain the M81 SF as well.

\subsection{Power Spectral Density}
\label{subsec:norm}
The remaining task is to determine $P_0$, the PSD normalization.
By definition, the variation power, $\int_{f_{1}}^{f_{2}} P(f) df$, 
is equivalent to RMS of the light curve
with bin width of 1/$f_2$ over the total observed span of 1/$f_1$.
Consequently,
we can relate $P_0$ to the RMS of the light curve,
when we specify $f_{\rm b}$ and $\alpha$.

In \S~\ref{subsec:Monte-Carlo}, 
step (iv) determines $P_0$ 
that corresponds to each 112-point light curve.
If there were no data gap, 
all light curves that are generated from the same $f_{\rm b}$ and $\alpha$
would have the same $P_0$.
In practice, however,
$P_0$ depends on the random numbers used in the Monte-Carlo simulation.
We accordingly made a histogram of 1000 values of $P_0$
assuming $1/f_{\rm b}=10^{4.6}$ days and $\alpha=1.40$,
and determined 90\% confidence region of $P_0$.
In reference to the best-estimate $P_0^{\rm best}$,
the 90\% confidence region in $P_0$ becomes
(0.8--1.2) $P_0^{\rm best}$.
Taking into account the uncertainty associated with $P_0$,
as well as $f_{\rm b}$ and $\alpha$ (Figure~\ref{fig:cont}a),
we have calculated a range of allowed PSDs 
with 68.3\%, 90\% and 99\% confidences.
In Figure~\ref{fig:cont}b,
we plot the PSD after dividing it by the square of its average flux;
the plot is called normalized PSD (NPSD) and 
convenient to compare random variation of different objects
(Miyamoto et al. 1994).

In connection with the normalization of the fake light curves,
we would like to stress one point.
In Figure~\ref{fig:SF},
the difference in the assumed $f_{\rm b}$ is
best reflected in the absolute level of SF
over the small-time-lag region;
shorter values of 1/$f_{\rm b}$ yield larger values of SF.
However this does not mean
that the information on longer time scales is unimportant.
In fact, the clear separation
among SFs with different PSDs occurs
because we have normalized the RMS of each fake light curve.
For the normalization to work properly, in turn,
we need to know the overall data variance.
When 1/$f_{\rm b}$ is longer than a few times the total span,
the averaged SFs with different values of 1/$f_{\rm b}$
are no longer discriminated from one another,
as shown in Figure~\ref{fig:cont}a.

\section{DISCUSSION}
\subsection{Comparison with Other Objects}
Let us compare the estimated PSD of the M81 nucleus
with those of other objects. 
We utilize the plot invented by Hayashida et al. (1998):
namely, a (frequency) $\times$ (NPSD) plot, or $fP(f)$ plot.
As Hayashida et al. (1998) pointed out,
the $fP(f)$ plot displays relative contribution of the power
at each frequency decade,
like the $\nu f_\nu$ plot of energy spectrum.
As illustrated in Figure~\ref{fig:npsd},
the power concentrates around $f_{\rm b}$,
so that $f_{\rm b}$ is an important parameter characterized each system.

So far, Ptak et al. (1998) and Awaki et al. (1999)
compared the high-frequency information of LLAGNs
with those of the other objects
and suggested that LLAGNs may have long time scale and/or large black-hole mass.
However, there was an implicit or explicit assumption in their analysis,
that the individual AGNs have the same PSD shape 
and same total area of NPSD on the $fP(f)$ plot.
There was no concrete evidence of the assumption,
and it should be examined
by measuring the low frequency portion of the PSD,
as we have performed.
Moreover, the slope index of PSD differs object to object ($1< \alpha< 2$).
Consequently, their result strongly depend on 
the frequency where the NPSD amplitude was measured.
Therefore, it is important to consider the PSD over wide frequency band.

In Figure~\ref{fig:npsd},
we show the M81 NPSD, derived from Figure~\ref{fig:cont}b, 
together with those of Seyfert galaxies and a Galactic black-hole candidate
taken from Hayashida et al. (1998) and Pounds et al. (1988).
The figure reveals two facts.
First,
the power peak of M81 locates at a very low frequency for its low luminosity.
Second,
the total area of the M81 NPSD is similar
to those of the other objects.
Consequently,
the observed lack of significant short-term variability in M81
is not because the total variation power is small,
but because the power concentrates on low frequencies.
This suggests that
the other LLAGNs with small NPSD amplitude at high frequency
also have large power on low frequencies.

\subsection{Black-Hole Mass}
Finally, we estimate the black-hole mass, $M_{\rm BH}$, of the M81 nucleus.
Conventionally, 
characteristic variation time scales of AGNs 
have been translated into $M_{\rm BH}$ 
assuming a linear proportionality between the time scale and mass,
and often utilizing Cygnus X-1 as a standard,
although this estimation is only appropriate
if the physics of accretion are similar between Cygnus X-1 and M81.
This allows us to estimate $M_{\rm BH}$ in the M81 nucleus to be
$> 4.0 \times 10^7~M_\odot$.

So far, $M_{\rm BH}$ of the M81 nucleus has been estimated in various ways.
From measurements of the velocity 
of the broad-line region or narrow-line region,
several authors estimated the mass within these regions of the M81 nucleus,
although their results have large scatter:
$1\times 10^6~M_\odot$--$1.8\times 10^7~M_\odot$ 
(Peimbert \& Torres-Peimbert 1981),
$(3-8)\times 10^5~M_\odot$ (Filippenko \& Sargent 1988),
and $7\times 10^5~M_\odot$--$3\times 10^6~M_\odot$ (Ho et al. 1996).
From stellar velocity dispersion measurements,
Keel (1989) obtained an upper limit of $2\times10^7 M_\odot$,
and Bower et al. (1996) obtained a lower limit of $3\times10^6 M_\odot$.
Our result falls on the relatively higher side of these previous estimates,
though still consistent 
considering large errors associated with various estimates.
Our estimate reinforces the view 
that LLAGNs have low X-ray luminosities
because they have very low accretion rates,
rather than because they have less massive black holes
than the more luminous AGNs.
This further encourages us 
to interpret LLAGNs as a relic of the past luminous AGNs.

\section*{Acknowledgments}

Thanks are due to 
Professor Hajime Inoue
for permission to use the {\it ASCA} data of 15th and 16th observations
and
Professor Kazuhisa Mitsuda 
for helpful discussions on several points in the paper.
This research was supported in part by a grant from 
Japan Society for the Promotion of Science.

\newpage

\noindent
{\large \bf References}\\

\begin{description}
\item Awaki H., Sakano M., Terashima Y., Hayashida K., 1999, AN, 320, 305
\item Barr P., Giommi P., Wamsteker W., Gilmozzi R., Mushotzky R., 1985, BAAS, 17, 608 
\item Boettcher M., Reuter H.-P., Lesch H., 1997 A\&A 326 33 
\item Boller Th., Meurs E.J.A., Brinkmann W., Fink H., Zimmermann U., Adorf H.-M., 1992, A\& A, 261, 57 
\item Bower G.A., Wilson A.S., Heckman T.M., Richstone D.O., 1996, in The Physics of Liners in view of recent observations. ASP Conference Series Vol.103, p.163 edited by M. Eracleous
\item Burke B.E., Mountain R.W., Harrioson D.C., Bautz M.W., Doty J.P., 
     Ricker G.R., Daniels P.J. 1991, IEEE Trans., ED-38 p1069
\item Edelson R., Nandra J., 1999, ApJ, 514, 682
\item Fabbiano G., 1988, ApJ, 325, 544 
\item Filippenko A.V., Sargent W.L.W., 1988, ApJ, 324, 134 
\item Freedman W.L., Hughes S.M., Madore B.F., Mould J.R., Lee M.G., Stetson P., Kennicutt R.C., Turner A. et al., 1994, ApJ, 427, 628 
\item Hayashida K., Miyamoto S., Kitamoto S., Negoro H., Inoue H., 1998, ApJ, 500, 642 
\item Heidt J., Wagner S.J., 1998, A\&A, 329, 853 
\item Ho L.C., Filippenko A.V., Sargent W.L.W., 1996, ApJ, 462, 183 
\item Ho L.C., Filippenko A.V., Sargent W.L.W., 1997, ApJ, 487, 568 
\item Hughes P.A., Aller H.D., Aller M.F., 1992, ApJ, 396, 469 
\item Ishisaki Y., Makishima K., Inoue H., Iyomoto N., Kohmura Y., Mitsuda K., Mushotzky R.F., Petre R. et al., 1996, PASJ, 48, 237 
\item Iyomoto N., 1999 PhD Thesis, University of Tokyo
\item Iyomoto N., Makishima K., 1999, AN, 320, 300
\item Iyomoto N., Makishima K., Fukazawa Y., Tashiro M., Ishisaki Y., Nakai N., Taniguchi Y., 1996, PASJ, 48, 231 
\item Iyomoto N., Makishima K., Matsushita K., Fukazawa Y., Tashiro M., Ohashi T., 1998, ApJ, 503, 168 
\item Kawaguchi T., Mineshige S., Umemura M., Turner E.L., 1998, ApJ, 504 
\item Keel W., 1989, AJ, 98, 195 
\item Kohmura Y., 1994, PhD Thesis, University of Tokyo  
\item Kohmura Y., Inoue H., Aoki T., Ishida M., Itoh M., Kotani T., Tanaka Y., Ishisaki Y. et al., 1994, PASJ, 46, 157 
\item Leighly K.M., O'brien P.T., 1997, ApJL, 481, L15 
\item Makishima K., Fujimoto R., Ishisaki Y., Kii T., Loewenstein M., Mushotzky R., Serlemitsos P., Sonobe T. et al., 1994, PASJ, 46, L77 
\item Makishima K. Tashiro M., Ebisawa K., Ezawa H., Fukazawa Y., Gunji S., Hirayama M., Idesawa E. et al. 1996 PASJ, 48, 171 
\item Miyamoto S., Kitamoto S., Iga S., Hayashida K., Terada K., 1994, ApJ, 435, 398 
\item Ohashi T., Ebisawa K., Fukazawa Y., Hiyoshi K., Horii M., Ikebe Y., Ikeda H., Inoue H. et al., 1996, PASJ, 48, 157 
\item Paltani S., Courvoisier T.J.-L., Blecha A., Bratschi P., 1997, A\&A, 327, 539 
\item Peimbert M., Torres-Peimbert S., 1981, ApJ, 245, 845 
\item Pellegrini S., Cappi M., Bassani L., Malaguti G., Palumbo G.G.C., Persic M., 2000, A\&A, 353, 447 
\item Petre R., Mushotzky R.F., Serlemitsos P.J., Jahoda K., Marshall F.E., 1993, ApJ, 418, 644 
\item Pounds K.A., McHardy I.M., 1988 in PHYSICS OF NUETRON STARS AND BLACK HOLES p.285 edited by Tanaka Y. Universal Academy Press  
\item Ptak A., Serlemitsos P., Yaqoob T., Mushotzky R., 1999, ApJS, 120, 179 
\item Ptak A., Yaqoob T., Mushotzky R.F., Serlemitsos P., Greffiths R., 1998, ApJL, 501, L37 
\item Serlemitsos P., Ptak A., Yaqoob T., 1996 in The Physics of Liners in view of recent observations. ASP Conference Series Vol.103, p.70 edited by M. Eracleous 
\item Simonetti J.H., Cordes J.M., Heeschen D.S., 1985, ApJ, 296, 46 
\item Terashima Y., Ptak A., Fujimoto R., Itoh M., Kunieda H., Makishima K., Serlemitsos P.J., 1998, ApJ, 496, 210 
\item Tsuru T., 1992, PhD Thesis, University of Tokyo 
\item Uno S., Mitsuda K., Inoue H., Takahashi T., Makino F., Makishima K., Ishisaki Y., Kohmura Y. et al., 1998, IAU Symopsium No. 188 P.245 
\item Wijnands R., van der Klis M., 1999, ApJ, 522, 965 
\end{description}

\newpage

\begin{figure}	
\caption{
The GIS (circle) and SIS (cross) fluxes of the M81 nucleus
in the 2--10 keV band,
obtained from spectral fits with a photon index fixed at 1.8.
Upper and lower panels correspond to the 1st through 6th 
and 7th through 16th observations, respectively.
Each bin represents each observation.
Systematic errors of the spectral fit are shown as error bars,
while Poisson errors are negligible.
\label{fig:longLC}
}
\end{figure}

\begin{figure}	
\caption{
The M81 short-term light curves with 1/8-day bin width.
Abscissa is time in days after 1993 March 23 (UT),
and ordinate is 2--10 keV flux in unit of erg s$^{-1}$ cm$^{-2}$.
Individual panels correspond to 
the light curves from the 1st through 16th observations.
Observation IDs are shown in the top left of each panel.
The vertical error bars correspond to the Poisson error.
The horizontal error bars represent the bin width of the light curve.
The $\chi^2$ and d.o.f. values against the constant-count hypothesis.
of the constant fit are shown
in the left bottom of each panel.
\label{fig:shortLC}
}
\end{figure}

\begin{figure}	
\caption{
(a) The raw M81 SF without binning,
made from the 1/8-day bin light curve (crosses).
For comparison,
small-time-lag portion of SFs 
made from 
1/4-day (triangles), 1/16-day (circles),
1/32-day (squares) and 1/64-day (diamonds) 
bin light curves are superposed.
(b) The M81 SF after binning.
The horizontal error bars represent the bin width of the SF.
(c) Examples of individual fake SFs,
having $\alpha=1.40$ and $1/f_{\rm b}=10^{4.0}$ days.
Each symbol corresponds each SF.
(d) Examples of ensemble averages of 1000 fake SFs, having $\alpha=1.40$ and 
$1/f_{\rm b}$=10 days (crosses), 100 days (circles), 
1000 days (squares) and 10000 days (triangles);
abscissa of the 100, 1000 and 10000 days SFs are shown with slight shifts
for convenience.
The standard deviations among 1000 Monte-Carlo simulations
are shown as vertical error bars.
(e) Ensemble averages of the fake SFs 
with various 1/$f_1$ of 
2$^{11}$ days (circles), 2$^{12}$ days (squares), 2$^{13}$ days (triangles), 
2$^{14}$ days (diamonds), 2$^{15}$ days (pluses), 2$^{16}$ days (asterisks), 
2$^{17}$ days (crosses) and 2$^{18}$ days (minuses).
The assumed PSD parameters are $\alpha=1.40$ and $1/f_{\rm b}=10^{4.6}$ days.
The standard deviations among fake SFs are similar to those in (d),
and are not shown for simplicity.
\label{fig:SF}
}
\end{figure}

\begin{figure}	
\caption{
(a) The $\Delta \chi^2$ distribution of 5000 fake light curves.
Here, $\Delta \chi^2 = \chi^2_{\rm true} - \chi^2_{\rm min}$.
(b) The $\chi^2_{\rm true}$ distribution
of 5000 fake light curves (histogram)
and $\chi^2_{\rm min}$ of actual M81 data (an arrow).
The M81 $\chi^2_{\rm min}$ lies within 90\% regions of the distribution,
confirming that the employed PSD shape is appropriate.
\label{fig:chi2}
}
\end{figure}

\begin{figure}	
\caption{
(a) Confidence contours for the M81 light curve
on the PSD index and break frequency plain.
A cross represents the parameters
that give the best fit to the {\it ASCA} light curve.
Dotted, solid and dashed lines represent 
68.3\%, 90\% and 99\% confidence regions, respectively.
(b) 
Estimated PSD of the M81 nucleus,
shown after normalized by the average flux.
Thick, medium and thin lines are 68.3\%, 90\% and 99\% confidence regions
of $f_{\rm b}$, $\alpha$ and $P_0$, respectively.
\label{fig:cont}
}
\end{figure}

\begin{figure}	
\caption{
Comparison between NPSD of the M81 nucleus and those of other objects,
in the form of frequency $\times$ NPSD plot.
Thick, medium and thin lines of the M81 NPSD
are 68.3\%, 90\% and 99\% confidence regions, respectively.
NPSDs of Cygnus X-1, MGC$-$6-30-15 and NGC 4051 and that of NGC 5506 
are taken from Hayashida et al. (1998) and Pounds et al. (1988), respectively.
\label{fig:npsd}
}
\end{figure}
\end{document}